\documentclass[12pt]{article}

\usepackage[squaren]{SIunits}
\usepackage{Fernandes_custom}
\usepackage{authblk}
\usepackage{cite}

\title{Optical force laws for guided light in linear media}
\author[1,2]{Thales Fernando Damasceno Fernandes}
\affil[1]{Universidade Federal de Minas Gerais (UFMG), Physics Department, 30123-970, Belo Horizonte, Minas Gerais, Brazil}
\affil[2]{Centro de Desenvolvimento da Tecnologia Nuclear – CDTN/CNEN, Av. Pres. Antônio Carlos, 6.627, Belo Horizonte, Minas Gerais, Brazil}
\author[3,*]{Pierre-Louis de Assis}
\affil[3]{``Gleb Wataghin'' Institute of Physics, University of Campinas – UNICAMP, Department of Applied Physics, 13083-859, Campinas, São Paulo, Brazil}
\affil[*]{plouis@ifi.unicamp.br}

\begin{document}

\flushbottom
\maketitle
\thispagestyle{empty}

\begin{abstract}	
	The mechanical response of transparent materials to optical forces is a topic that concerns a wide range of fields, from the manipulation of biological material by optical tweezers to the design of nano-optomechanical systems (NOMS).
	However, the fundamental aspects of such forces have always been surrounded by controversies, and several different formulations have been proposed. In this work, we focus on the specific case of light propagating as a superposition of guided modes in lossless dielectric waveguides as a physical example upon which to build a general stress tensor.
	We use this formalism to calculate optical forces for straight and curved waveguide sections and all possible excitation configurations for a given set of coupled eigenmodes, and then compare the results for each of the known proposed optical force laws as well as a novel one derived from this general stress tensor.
	We show that proper use of the divergence theorem is crucial to account for all force terms, many of which vanish if the procedure most commonly used is applied for situations other than eigenmodes in straight waveguides.
	A better understanding of how different stress tensors predict very different forces for certain waveguide geometries opens a pathway for new experimental tests of each formulation.
\end{abstract}

\section*{Introduction}
It has been known since Maxwell that electromagnetic waves carry momentum and can exert forces on material objects when reflected, refracted or absorbed~\cite{Abraham1909}.
Mechanical interactions between light and matter have been developed into applications that range from atomic cooling~\cite{Phillips1982} to micro-optomechanical devices and the manipulation of living cells using optical tweezers~\cite{Ashkin1987,Chiou2005}.

For all the applications already developed, however, there are still points in the theory of optical forces that are open to discussion.
The most well-known is the Abraham-Minkowski controversy about how to properly define the momentum of light in dielectric media~\cite{Mansuripur2010a,Ramos2011,Kemp2011}.
Less known is the multitude of proposed force laws or methods to calculate the optical force~\cite{Mansuripur2008,Zakharian2006,Mansuripur2013a,Jazayeri2014,Rubinsztein-Dunlop2017}.
These may disagree in certain situations while agreeing in others~\cite{Kemp2011}.

In order to compare the different proposed optical force laws in a way that allowed us to discern the reasons for their disagreements in certain conditions and agreements in others, we have chosen to focus our study on a simple, yet enlightening, system composed of a pair of non-magnetic linear lossless dielectric waveguides evanescently coupled.
By studying the stationary regime of light propagation and considering static waveguides, we were able to isolate the optical force analysis from the context of more complex phenomena such as photon-phonon interactions~\cite{Li2014,VanLaer2015}, especially Brillouin scattering~\cite{Ippen1972,Shin2013}.
While the forces studied in our work may provide an interaction pathway for Brillouin scattering in certain photonic devices, that is not within the scope of our investigation.
Our choice of model constraints has also set the optical forces under study apart from electrostrictive effects which are intrinsically dependent on deformations of the dielectric material.

In the following sections, we present a way to construct a general tensor formulation that can be used to recover all the possible force laws in dielectrics by a choice of two binary parameters, followed by numerical calculations.
Our calculations used a pair of coupled waveguides of rectangular cross-section that supported two TE-like and two TM-like modes.
In addition to varying their separation during simulations, we have also considered the special case of zero gaps as a single waveguide of double width.
We show that even in this case there are important differences to be considered between force laws.
Due to the form of the general optical force derived from our general stress tensor, we have simulated not only straight sections but also circular sectors, since some terms might vanish for straight waveguides but not for those with a curvature.
We compared the case of single eigenmode propagation to that of a superposition of the first two eigenmodes with same polarization for coupled waveguides and their single-waveguide limit, in both the straight and curved geometries.

\section*{General Tensor Formulation}
The momentum balance in an electromagnetic system can be cast in a simple form as~\cite{Jackson1998}:
\begin{align}
	\vc{F} + \Dps{\vc G}{t} &= \Div{\vc T}, \label{eq:MomentumBalance}
\end{align}
where $t$ is the time, $\vc{F}$ is the force density, $\vc{G}$ the momentum density, and $\vc{T}$ the stress tensor.

Let $\vc{T}\!_{nm}$ be the general stress tensor defined as:
\begin{align}
	\vc{T}\!_{nm} &= \varepsilon_0 \Bracket{
		\varepsilon^n \TensorProduct{\vc E}{\vc E} - \frac{\id}{2} \varepsilon^m \Dot{\vc E}{\vc E}
	} + \mu_0 \Bracket{
		\TensorProduct{\vc H}{\vc H} - \frac{\id}{2} \Dot{\vc H}{\vc H}
	}, \label{eq:StressTensor}
\end{align}
where $n$ and $m$ can be either $0$ or $1$ and will be used to label different stress tensors; $\vc E$ and $\vc H$ are the time-dependent electric and magnetic fields, respectively; $\varepsilon$ is the relative permittivity of the material, and $\id$ is the identity matrix.
Henceforth it will be assumed the material is a non-magnetic linear lossless dielectric and that the guided waves are monochromatic.
With this general definition, we have that
$\vc{T}\!_{00}$ is the Lorentz stress tensor~\cite{Rubinsztein-Dunlop2017,Jazayeri2014},
$\vc{T}\!_{10}$ is the Einstein-Laub stress tensor~\cite{Jazayeri2014},
$\vc{T}\!_{11}$ is the Minkowski stress tensor~\cite{Jazayeri2014} and
$\vc{T}\!_{01}$ is not found in the literature, to the best of our knowledge.

Taking the divergence of the general stress tensor and using the momentum balance Eq.~\eqref{eq:MomentumBalance} we can define the following force densities~\cite{Jazayeri2014,Rubinsztein-Dunlop2017}:
\begin{align}
	\vc{F}_{00} &= -\vc{E} \Div{\vc P} + \mu_0 \Cross{\Dps{\vc P}{t}}{\vc H}, && \text{Lorentz} \label{eq:ForceLorentz} \\
	\vc{F}_{10} &= \DotNabla{\vc P}{\vc E} + \mu_0 \Cross{\Dps{\vc P}{t}}{\vc H}, && \text{Einstein-Laub} \label{eq:ForceEinstein-Laub} \\
	\vc{F}_{11} &= -\frac{1}{2}\varepsilon_0\Dot{\vc E}{\vc E}\Grad{\varepsilon},  && \text{Minkowski} \label{eq:ForceMinkowski} \\
	\vc{F}_{01} &= \vc{F}_{00} + \vc{F}_{11} - \vc{F}_{10}, && \text{Unnamed (Previously unreported)} \label{eq:ForceUnnamed}
\end{align}
where $\vc P = \varepsilon_0(\varepsilon-1)\vc{E}$ is the polarization density.

Taking the time-average of the momentum balance Eq.~\eqref{eq:MomentumBalance}, we have:
\begin{align}
	\Mean{\vc{F}} + \underbrace{\Mean{\Dps{\vc G}{t}}}_{=0} &= \Div{\Mean{\vc T}}, \label{eq:meanF}
\end{align}
where $\Mean{\Dps{\vc G}{t}}=0$ comes from the time-averaging process of time-harmonic fields.
Hence, for time-harmonic fields, which we were concerned in this study, terms originated from the electromagnetic momentum will not contribute to the time-averaged force $\Mean{\vc{F}}$.

The force densities shown above are very different in form, even though they originate from quite similar stress tensors.
While a comparison of their effects based solely on the examination of the formulations presented in Eq.~\ref{eq:ForceLorentz} to~\ref{eq:ForceUnnamed} is a complex task, we can indeed analyze their differences by using a model system composed of dielectric waveguides that are assumed to be perfectly rigid.
This condition is reasonable for all time scales longer than the optical periods, typically in the femtoseconds, but still much shorter than those associated to acoustic responses of solids, which are no shorter than some nanoseconds.
This allows us to examine the optical forces in a context that is free of other optomechanical effects such as electrostriction or photon-phonon scattering, and study the regime of statical deformations.

In order to facilitate our calculations, we used the standard procedure of applying the divergence theorem to cast the total force in terms of a surface integral instead of the volume integral of a divergence.
This avoids field derivatives altogether, which is particularly beneficial when using the Finite Elements Method (FEM) in numerical simulations.
The total force can be given by
\begin{align}
\Vec{F}_\mathrm{tot} &= \iiint_\Omega \Div{\Mean{\vc T}} \dd{V} 
= \iint_\Gamma  \Dot{\Mean{\vc{T}_\mathrm{out}}}{\Vec[unit]{n}}\, \dd{S}, \label{eq:Ftot}
\end{align}
where $\Omega$ is the volume of integration, $\Gamma$ is its boundary, and $\Vec[unit]{n}$ is the outward normal unit vector.
When using the divergence theorem for electromagnetic fields in dielectrics, one needs to take into account the discontinuity of the fields at the boundary and use the proper discontinuous form of the divergence theorem~\cite{Negahban2005,Khoei2015}.
It can be shown that the fields to be taken into account in a discontinuous stress tensor are the fields outside the domain of integration ($\vc{T}_\mathrm{out}$)~\cite{Kemp2005}.

\begin{figure}
	\centering
	\includegraphics[width=\linewidth]{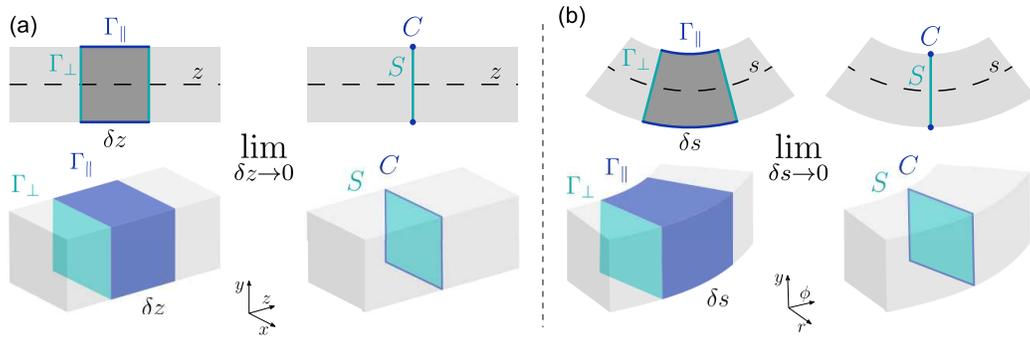}
	\caption{
		Surface of integration for Eq.~\eqref{eq:Ftot}.
		In (\textbf{a}) is shown the simple case of propagation in the $z$-axis, whereas in (\textbf{b}) is shown the case a circular path in cylindrical coordinates, parametrized by the arc-length $s$. Although a particular case, it can be used to approximate infinitesimal sections of most arbitrary curved paths.
		$\Gamma_\parallel$ defines a surface parallel to the propagation direction, whereas $\Gamma_\perp$ is a perpendicular one.
		In the limit where the thickness $\delta s$ or $\delta z$ of the bounded volume tends to zero, the surface $\Gamma_\parallel$ is transformed in the line $C$, whereas the surface $\Gamma_\perp$ becomes the surface $S$.
		The first row shows the top-view of the waveguides whereas the second row shows the perspective view of the waveguides.
	}
	\label{fig:int-surface}
\end{figure}

In our model system of light propagating in coupled waveguides with rectangular cross-sections, the boundary $\Gamma$ can be divided into two parts: one perpendicular, $\Gamma_\perp$, and another one parallel to the direction of propagation, $\Gamma_\parallel$, illustrated in Fig.~\ref{fig:int-surface}.

For applications in which a static structure deformation profile is to be calculated, the knowledge of the total force is not enough, and the linear force density along the propagation direction, $\Dps{\Vec{F}}{s}$, is also required.
We parametrized the propagation direction by an arc length $s$ so that curved waveguides could also be described with ease.
To calculate $\Dps{\Vec{F}}{s}$ we will integrate Eq.~\ref{eq:Ftot} on a slice of infinitesimal thickness $\delta s$ in the $s$-axis and take the limit $\delta s\rightarrow0$, as shown in Fig.~\ref{fig:int-surface}(b).
In this limit, the surfaces $\Gamma_\perp$ and $\Gamma_\parallel$ become the surface $S$ and the line $C$, respectively, illustrated in Fig.~\ref{fig:int-surface}.
The force density on that slice becomes
\begin{align}
	\Dps{\Vec{F}}{s} &= 
	{\iint_S \Dot{\Mean{\Dps{\vc{T}_\mathrm{in}}{s}}}{\Vec[unit]{n}} \,\dd{S}}
	+
	{ \iint_S \Dot{\Mean{\vc{T}_\mathrm{in}}}{\Dps{\Vec[unit]{n}}{s}} \,\dd{S}}
	+
	{ \oint_C \Dot{\Mean{\vc{T}_\mathrm{out}}}{\Vec[unit]{n}} \,\dd{l}}, \label{eq:dF/ds}
\end{align}
where the first and second terms are integrated over the cross-section area $S$ and the third term is a line integral over the boundary of $C$.
In the surface $S$, the fields outside the domain of integration are the same as the fields inside it, since they are continuous in this surface, thus, we renamed the tensor in these cases as $\vc{T}_\mathrm{in}$ to represent the fields inside matter to avoid confusion.

For eigenmodes propagating in the $z$-axis, Fig.~\ref{fig:int-surface}(a), we have $\Dps{\Vec[unit]{n}}{z}=0$, since the normal in the cross-section is not changing, and $\Mean{\Dps{\vc{T}_\mathrm{in}}{z}}=0$, since for eigenmodes the stress tensor is $z$-independent.
In this specific case of eigenmodes propagating in a section of zero curvature, only the third integral in Eq.~\eqref{eq:dF/ds} survives and the force density $\Dps{\Vec{F}}{z}$  will be formulation-independent, depending solely on the fields outside the dielectric.
It is important to notice that we assume that $\varepsilon=1$ outside the material, a situation in which  Eq.~\eqref{eq:StressTensor} shows that all stress tensors will coincide with the Lorentz formulation.
However, if the waveguides are surrounded by another dielectric material, different formulations will lead to different predictions even for eigenmode forces.

Since the first and second term in Eq.~\eqref{eq:dF/ds} depend on the fields inside the material ($\vc{T}_\mathrm{in}$), in a general scenario different force laws will result in different forces, as we have shown in a previous work, when studying the appearance of a beating force due to non-eigenmode excitation~\cite{Fernandes2018}.
In summary, a non-eigenmode excitation will have a non-null first term in Eq.~\eqref{eq:dF/ds}, whereas a curved propagation will have a non-null second term due to the varying normal.
The third term in Eq.~\eqref{eq:dF/ds} will always be present, but on certain occasions it can be zero due to symmetries.

\section*{Results and Discussion}
In order to evaluate quantitatively the behaviors discussed in the previous section, we have run simulations using the Finite Element Methods (FEM) for silicon waveguides with a refractive index of $3.45$ at an excitation wavelength of $1550$~nm.
We modeled waveguides using a rectangular cross-section with a width of $280$~nm and height of $380$~nm.
For the single waveguide geometry (zero gaps) the width was $560$~nm, while the height remained $380$~nm.
Curved sections were simulated using a radius of curvature of $4~\upmu$m.

\begin{figure}
	\centering
	\includegraphics[width=\linewidth]{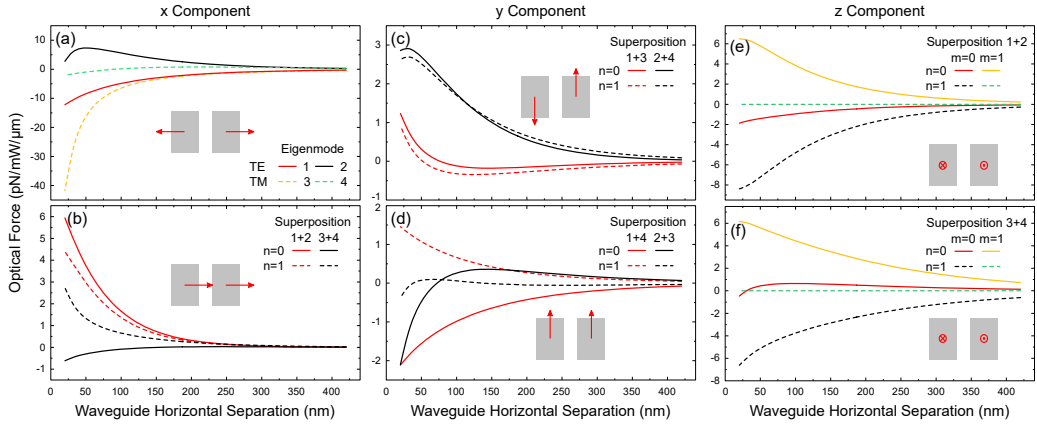} 
	\caption{
		Comparison of optical forces calculated using Eq.~\eqref{eq:dF/ds} for a pair of straight coupled waveguides.
		In (\textbf{a}) are shown the eigenmode forces, in the transverse $x$-direction, for the $4$ modes in this system, where two are TE and two are TM.
		Since all force models agree for eigenmodes, a single line is presented for each mode.
		For mode superpositions, only the beating force components are shown in (\textbf{b-f}).
		Given the symmetries of each mode, only some cartesian components of the optical beating forces will be non-zero for each mode superposition.
		The $x$ component of the beating force, shown in (\textbf{b}), is present for mode superpositions $1+2$ and $3+4$.
		It can be seen that the force has the same sign on both waveguides and therefore acts on the center of mass of the cross-section.
		The $y$ component presented in (\textbf{c,~d}) is non-zero for mode superpositions $1+3$, $2+4$, $1+4$, and $2+3$.
		It can be seen in (\textbf{c}) that for superpositions $1+3$ and $2+4$, the $y$ component acts as a shear force, orthogonal to both the gap between waveguides and the direction of propagation.
		Conversely, in  (\textbf{d}) it can be seen that for $1+4$ and $2+3$ the $y$ component of the force acts on the center of mass of the cross-section.
		This contrast is due to the $y$ parity of modes $1$ and $3$, as well as $2$ and $4$, being the same~\cite{Fernandes2018}. In (\textbf{e}, \textbf{f}) it is shown that the longitudinal component, $z$, of the beating forces only appears for superpositions $1+ 2$ and $3+4$, acting as a shear force in both cases.
		For (\textbf{b}-\textbf{d}), only the $n$ value is relevant of Eq.~\eqref{eq:StressTensor}, and the forces split in two groups of similar qualitative behavior, whereas in (\textbf{e}, \textbf{f}), there are $4$ different longitudinal components, due to the dependence on both $n$ and $m$.}
	\label{fig:forces}
\end{figure}

Figure~\ref{fig:forces} shows the optical force densities calculated using Eq.~\eqref{eq:dF/ds} for a pair of straight coupled waveguides, normalized for input power, as a function of their separation.
They were excited by eigenmodes and by a superposition of eigenmodes.
For the case of the superposition, only the additional beating force term is shown (first term in eq.~\eqref{eq:dF/ds}), in a similar fashion to our previous result~\cite{Fernandes2018}.
Fig.~\ref{fig:forces}(a) shows the forces for the $4$ eigenmodes on the system.
Figs.~\ref{fig:forces}(b-f) show the beating force on the system due to the superposition of eigenmodes. Figs.~\ref{fig:forces}(b-d) show that the combination of different symmetries yields forces in different directions as well as with different characteristics.
Figs.~\ref{fig:forces}(b, d) show a beating force that acts on the center of mass of the cross-section, whereas Figs.~\ref{fig:forces}(c, e, f) show a beating force that acts as a shear force (opposite sign on opposite waveguides, but orthogonal to the gap between them). 

Turning to the comparison between optical force models, Fig.~\ref{fig:forces}(a) also shows the important fact that for eigenmodes all force laws give the same result, as already proved.
Therefore, a single curve is presented for each mode.
For the beating force, transverse components ($x$ and $y$), Figs.~\ref{fig:forces}(b-d), depend only on the value of $n$ of Eq.~\eqref{eq:dF/ds}, a fact that can be verified by expanding the equations.
It should be noted that the transverse components along the $y$-direction are weaker than those along the $x$-direction due to being orthogonal to the gap between waveguides.
The longitudinal component of the beating forces ($z$), Figs.~\ref{fig:forces}(e, f), depends on both values of $n$ and $m$. 

In the case of $1+3$ and $2+4$ superpositions, yielding a $y$ shear force, shown in Fig.~\ref{fig:forces}(c), the forces predicted are very similar for different values of $n$.
Differences are more marked for the $3+4$ (black curves in Fig. ~\ref{fig:forces}(b)), $1+4$, and $2+3$ (Fig. ~\ref{fig:forces}(d)) mode superpositions, where some qualitative differences can be seen as well as quantitative ones.
The most striking result, however, is that of the $z$ component for superpositions $1+2$ and $3+4$, where the Einstein-Laub $(n = 1, m = 0)$ and Unnamed $(n = 0, m = 1)$ models predict forces with opposite sign.
Given their high magnitude at $100$~nm separation, which is well within reach of current nanofabrication technology, this component is a good candidate for experimental tests of optical force models, the main challenge being the optimization of a device to measure a shear force in the longitudinal direction.

\begin{figure}
	\centering
	\includegraphics[width=\linewidth]{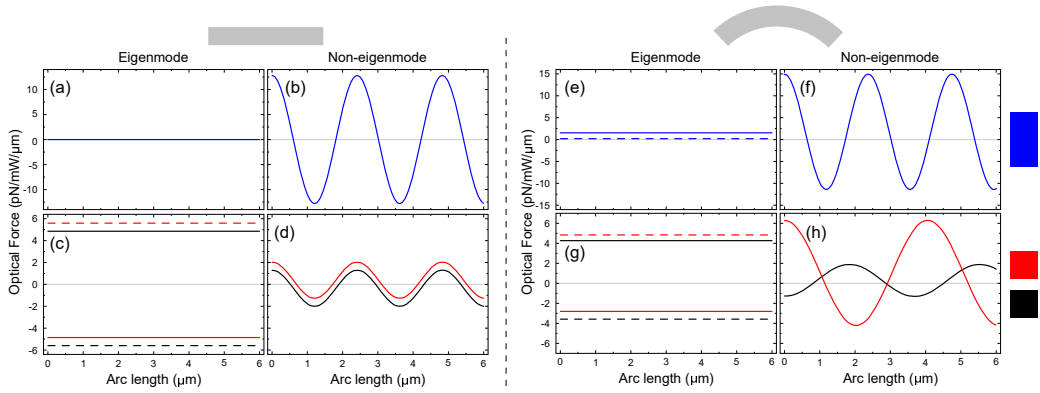}
	\caption{
		Transverse optical forces ($x$-direction) variation along the propagation direction for single and a pair of waveguides, curved and straight.
		In (\textbf{a},~\textbf{b},~\textbf{e},~\textbf{f}) are show the case of a single waveguides, whereas in (\textbf{c},~\textbf{d},~\textbf{g},~\textbf{h}) are show the case of a pair of coupled waveguides.
		In (\textbf{a},~\textbf{c},~\textbf{e},~\textbf{g}) are show only eigenmode forces, whereas in (\textbf{b},~\textbf{d},~\textbf{f},~\textbf{h}) are show the total force in the superposition (eigenmode forces + beating force).
		Blue curves represent forces of only one waveguide, whereas red and black curves represent the force on the top/bottom waveguide, respectively.
		The dashed curves represent the force of the second mode used in the superposition.
		Here is only shown the Lorentz force.
	}
	\label{fig:path}
\end{figure}

When analyzing a curved section, we have found that all values of $n$ and $m$ would give different forces.
Therefore using the same approach used for straight sections would not be fruitful in terms of comparisons, as the number of optical force densities would grow drastically (there would be $16$ forces for eigenmodes alone instead of $4$).
Given this difficulty, we chose to adopt a different approach to obtain information about the qualitative behavior of the optical forces in curved geometries.
Instead, Fig.~\ref{fig:path} shows the optical force for single (a,~c,~e,~g) and double (b,~d,~f,~h) waveguides, and for straight (a-d) and curved (e-h) waveguides.
Due to the symmetry of the fields, for a single waveguide, the eigenmode force is zero (a), whereas, for a curved waveguide (e), there is a force appearing due to the bending, and thus, changing linear momentum.
An interesting thing occurs with the superposition for a pair of straight waveguides, (d), the beating force gets summed to the eigenmode forces, thus making the forces on each waveguide different.
The space between the curves can get shortened (when the eigenmode forces cancel) or can grow apart.
The modulation of the curve can be changed by the amount of each mode in the superposition.
Thus, it is possible to control precisely the forces on each waveguide.
A similar feature occurs if the waveguides are curved, (h), but in this case, the forces have opposite signals in opposite waveguides (but they do not cancel each other), and also, they have different beating lengths.
It can easily seem from these examples that if the proper method to calculate the forces were not employed all these features would be lost.
In this particular case the Lorentz force was used to calculate all forces, if a different force law were used, different magnitudes and signals would be found, but the general beating behavior would still remain.

\section*{Conclusion}
Even though classical electromagnetism has been under study for more than a century, some core issues remain open. One such issue is that of the correct force law to describe the action of optical fields on dielectric materials. In order to further the study of this topic, we have started by casting the possible formulations of optical forces in lossless non-magnetic dielectrics in terms of a generic optical stress tensor. By doing so, we have not only created a common ground for comparison between the Lorentz, Einstein-Laub, and Minkowski formulations but also obtained a fourth one, which can be written as a linear combination of the previously known tensors, and with no evident physical motivation at present.

As we have shown, when departing from the highly symmetrical case of eigenmodes propagating along infinite parallel waveguides towards either superposition of eigenmodes or curved geometries, great care must be taken with the calculation of the optical forces. Boundary surfaces that do not contribute in the former case are highly significant to the latter, due to the breaking of invariance along the direction of propagation. This is a fact that has essentially gone unnoticed in the literature, given the emphasis on the study of eigenmode forces. We would like to stress that by putting all formulations in the same tensor framework we were able to clearly distinguish the new contributions that arise from having eigenmode superpositions from those that arise from a curvature of the guiding structure, even when a single eigenmode is studied. Such an analysis would not be possible by taking each force law in its particular form.

Finally, we believe that this discussion should be followed by experimental tests since in some situations the predicted result is markedly different for each force law. In addition to contributing to the ongoing discussion about optical forces in dielectrics, we believe that our contribution can also find applications in the design of integrated optomechanical devices, much in the same way as the inclusion of additional material effects in the modeling has lead to the design of new devices based on high photoelastic coupling instead of the so-called moving boundary coupling due to radiation pressure.

\end{document}